\documentclass[aps,onecolumn,12pt,preprintnumbers]{revtex4}

\begin{document}
 
\newcommand{\dfrac}[2]{\frac{\displaystyle #1}{\displaystyle #2}}
\preprint{VPI--IPPAP--04--02}
 
\title{Some Consequences of the Hypothesis of Minimal Lengths}
\author{Lay Nam Chang}\email{laynam@vt.edu}
\affiliation{Institute for Particle Physics and Astrophysics,
Physics Department 0435, Virginia Tech, Blacksburg VA 24061}

\begin{abstract}
\begin{center}
Talk presented at the 3rd International Symposium on Quantum Theory 
and Symmetries, September 13, 2003, University of Cincinnati, OH.
\end{center}
\end{abstract}

\maketitle

\begin{center}
\textit{Dedicated to the memory of Freydoon Mansouri}
\end{center}


Freydoon Mansouri and I first met during the fall of 1970.  
I was in the second year of my postdoc at the Enrico Fermi Institute 
at the University of Chicago, and Freydoon was beginning his first 
postdoc there just after graduating from Johns Hopkins.  
As it turned out, I had been studying his work with Gabor Domokos, 
and when I was given the opportunity to share my office with him, 
I promptly accepted the offer.  
Thus began over thirty years of friendship, support, 
intellectual give-and-take, and warm memories.  
And a satisfying collaboration that lasted for almost ten years.  
But even though we went on to work with different collaborators, 
we both knew that we had established a lasting relationship, 
and that we could count on each other in times of problems both big and small.  
I shall miss him.

It is my honor to dedicate this essay to Freydoon Mansouri.  

The subject of this paper is in many ways an outgrowth of what 
Freydoon and I did as our first joint research.  
We were examining the dynamical symmetries underlying duality, 
and in particular, to explore the extent to which strings could be 
the proper variables to use in embodying these symmetries.  
The basic entities in strings are extended structures, not points, 
and the position variables describing these structures do not necessarily 
commute. So the geometry underlying physical systems is non-commutative.  
I will be describing to you results that have been obtained in work carried 
out jointly with Djordje Minic, Tatsu Takeuchi, both faculty members at 
Virginia Tech, Naotoshi Okamura, a postdoc, and with Sandor Benczik and 
Saifuddin Rayyan, both graduate students in our group.  

The precise nature of the geometry is uncertain, 
but there are clues from a perturbative treatment of strings.  
For distances large compared to the length scale implied by the string tension, 
excitations of the system could be described by positions and momenta, 
at least for energies for which pair creations are expected to be negligible.   
Perturbative treatments of strings suggest that, 
concomitant with the non-commutativity in position,  
the uncertainties in position and momenta are modified to \cite{MLUR}:
\begin{equation}
\Delta x \geq {\hbar\over 2}\left({1\over \Delta p} + \beta \Delta p \right)\;,
\end{equation}
which implies
\begin{equation}
\Delta x \geq \hbar \sqrt{\beta}\;,
\end{equation}
so a minimal length is embedded naturally in the system.  
Note the UV/IR mixing in the uncertainty relation, 
giving rise to possibilities of effects of short distance phenomena 
to be present at lower energies.

While the perturbative treatments cannot fix the precise commutation 
relations among canonical variables, the uncertainty relations suggest 
some simple forms for these relations.  In what follows, we posit relations 
for the variables describing the positions and momenta of point particles 
as an abstraction of the underlying string dynamics which are consistent 
with these uncertainty relations \cite{kempf}:
\begin{eqnarray}
[\, \hat{x}_i,\, \hat{p}_j \,] & = & 
i\hbar\,(\, \delta_{ij} + \beta  \hat{p}^2 \delta_{ij}
                    + \beta' \hat{p}_i \hat{p}_j 
      \,)\;,\cr
[\, \hat{p}_i,\, \hat{p}_j \,] & = & 0 \;,\cr
[\, \hat{x}_i,\, \hat{x}_j \,] & = & i\hbar\,
  \frac{(2\beta-\beta') + (2\beta+\beta')\beta\hat{p}^2}
       { (1+\beta \hat{p}^2) }
  \left( \hat{p}_i \hat{x}_j - \hat{p}_j \hat{x}_i
  \right) \;,\cr
\hat{L}_{ij} & = & \frac{ \hat{x}_i \hat{p}_j - \hat{x}_j \hat{p}_i }
                        { (1+\beta \hat{p}^2) } \;.
\end{eqnarray}
The parameters $\beta$ and $\beta'$ are relics of the string tension, 
and should be thought of as the expansion coefficients of expressions 
that are derivable from string theory.  We have chosen to modify the 
canonical relationships first, as an expansion, and with the supposition 
that momenta commute, the non-commutativity of the position operators 
can then be deduced by appeal to the Jacobi identity.  
The relations are supposed to hold in $D$-dimensions.  
It can readily be checked that the operators $\hat{L}_{ij}$ satisfy 
the algebra of angular momentum. 

An underlying assumption we shall be making in what follows is that 
the correspondence principle applies, and that the dynamical systems 
continue to be controlled by Hamiltonians that retain their forms 
unchanged from the limits $\beta=0$ and $\beta'=0$.  
With these caveats, let us first examine how the classical canonical 
structure gets modified.  For definiteness, we will define the classical 
limit by replacing commutators with the corresponding Poisson brackets:
\begin{equation}
\left\{F, G \right\} = 
\left( \frac{\partial F}{\partial x_i} \frac{\partial G}{\partial p_j}
     - \frac{\partial F}{\partial p_i} \frac{\partial G}{\partial x_j}
\right) \left\{ x_i , p_j \right\} 
+ \frac{\partial F}{\partial x_i}\frac{\partial G}{\partial x_j}
\left\{x_i, x_j\right\} \;.
\end{equation}
The modified canonical equations of motion now take the form
\begin{eqnarray}
\dot{x}_i & = & \left\{ x_i, H \right\} 
\;=\; \phantom{-}
      \left\{ x_i , p_j \right\} \frac{\partial H}{\partial p_j}
    + \left\{ x_i , x_j \right\} \frac{\partial H}{\partial x_j} \;,\cr
\dot{p}_i & = & \left\{ p_i, H \right\}
\;=\; - \left\{ x_i, p_j \right\} \frac{\partial H}{\partial x_j} \;,
\end{eqnarray}
and Newton's force equations become
\begin{eqnarray}
m\ddot{x_i} & = & (1+\beta p^2)\,(1+3\beta p^2)\, F_i \;, \cr
F_i & \equiv & -{\partial V \over \partial x_i } \;.
\end{eqnarray}
For illustrative purposes, we set $\beta' = 0$ in the above.  
For the case of the $1/r^2$ gravitational force, the equation points 
to a breakdown of the equivalence principle \cite{rotationcurve}.  

To get a feel for what is happening, let us specialize to the case of 
central potentials, with Hamiltonians given by
\begin{equation}
H = {p^2 \over 2m} + V\left(r\right) \;,
\end{equation}
where the potential has zero Poisson brackets with $L_{ij}$.  Specifically,
\begin{eqnarray}
\dot{x}_i 
& = &  \left[\, 1 + (\beta + \beta')p^2 \,\right]\frac{p_i}{m} \cr
&   & +\left[\,(2\beta - \beta') + (2\beta + \beta')\beta p^2 \,\right]\frac{F}{r}\,L_{ij}\,x_j \;,\cr
\dot{p}_i 
& = & \left[\, (1+\beta p^2) x_i +\beta' (p\cdot x) p_i \,\right]\frac{F}{r}\;,\cr
F & \equiv & -\frac{\partial V}{\partial r}\;,
\end{eqnarray}
with 
\begin{eqnarray}
\left\{ x_k , L_{ij}\right\} & = & x_i \delta_{jk} - x_j \delta_{ki} \;,\cr
\left\{ p_k , L_{ij}\right\} & = & p_i \delta_{jk} - p_j \delta_{ki} \;.
\end{eqnarray}
As in the normal case, motion occurs in a plane.  But, for neither the simple 
harmonic potential, nor for the Coulomb potential are the orbits closed \cite{mercury}.
For simple harmonic motion, for the simpler case of $\beta'=0$, the resultant 
precession is an advancement, contrary to the normal case, while for the Coulomb 
force, it is a retardation.  We may attempt to place a limit on $\beta$ through 
its effect on the orbit of Mercury coming solely from this retardation:
\begin{equation}
\hbar\sqrt{\beta} < 2.3 \times 10^{-68} {\rm m}\;.
\end{equation}
This limit is intended for illustrative purposes only, 
since modifications on other influences on this orbit have not been included.  

The equations of motion do yield a modified Liouville theorem.  
It can be checked that the following is left invariant during evolution \cite{cosmoconst}:
\begin{equation}
\frac{ d^D \mathbf{x}\, d^D \mathbf{p} }
     { \left[ 1 + \beta p^2 \right]^{D-1}
       \left[ 1 + \left(\beta + \beta'\right) p^2 \right]^{1-\beta'/2(\beta+\beta')}
     }\;.
\end{equation}
Using this volume, we can estimate the corrections to the zero point energy 
coming from the existence of a minimal length \cite{cosmoconst}.  
For $\beta' =0$, and $D=3$,
\begin{eqnarray}
\Lambda \left(m\right) 
& = & \int \frac{d^3\mathbf{p}}{(1+\beta p^2)^3} \left(\frac{1}{2}\sqrt{p^2 + m^2} \right)\cr
& = &  2\pi \int_0 ^\infty \frac{p^2 dp}{(1 +\beta p^2)^3} \sqrt{p^2 + m^2} \cr
& = & \frac{\pi}{2\beta^2} \,f\!\left( \beta m^2 \right)\;,
\end{eqnarray}
\begin{equation}
f(x) = \left\{
\begin{array}{ll}
1 + \dfrac{x}{2(1-x)} + \dfrac{x^2}{4(1-x)^{3/2}}
    \ln\left( \dfrac{1 - \sqrt{1-x}}{1 + \sqrt{1-x}} \right) 
& \qquad (x\le 1)\;, \\
1 - \dfrac{x}{2(x-1)} + \dfrac{x^2}{2(x-1)^{3/2}} \tan ^{-1} \sqrt{x-1}
& \qquad (x\ge 1)\;.
\end{array}
\right.
\end{equation}
Asymptotically,
\begin{eqnarray}
f(x) & \approx & 
\left\{\begin{array}{ll} \sqrt{x} & \qquad \mbox{for large $x$}\;, \\
                        \left(1+x\right)^{0.42} & \qquad  0 < x < 1\;,
       \end{array} 
\right. \cr
\Lambda (0) & = & \frac{\pi}{2\beta ^2}\;.
\end{eqnarray}
As expected, the putative cosmological constant is finite, 
with higher frequencies being quenched by the minimal length.  
The corresponding effects on blackbody spectra are however quite slight.  

We now examine some quantum consequences of the commutation relations posited above.  
We will first examine what happens when $D=1$.   For the sake of concreteness, 
we will examine the simple harmonic oscillator in 1-dimension.  
In many respects this example is trivial, since the deformed relations 
can always be recast into the standard form, as we will demonstrate below.   
Since we are supposing that momentum variables remain commutative, 
we will work in momentum space throughout.  
In which case, the following representation reproduces the modified canonical 
relationship between position and momentum:
\begin{eqnarray}
\hat x & = & i \hbar\left\{\left(1 + \beta p^2\right) {\partial \over \partial p} + \gamma p \right\}, \cr
\hat p & = & p \;.
\end{eqnarray}
Hermiticity requires
\begin{eqnarray}
\langle f | g \rangle & = & \int {dp \over \left(1 + \beta p^2\right)^{1-\alpha}} f^\ast (p) g (p)\;, \cr
\alpha & = & \frac{\gamma}{\beta}\;.
\end{eqnarray}
The Hamiltonian is taken to be
\begin{equation}
\hat{H} = \frac{1}{2}m\omega^2 \hat{x}^2 + \frac{1}{2m}\hat{p}^2\;.
\end{equation}
The resulting Schr\"odinger equation cannot be easily compared with standard forms.  
However, it is straightforward to perform a change of variables so 
that it can be so compared.  We define
\begin{equation}
\rho = \frac{1}{\sqrt{\beta}} \tan ^{-1} \left(\sqrt{\beta} p \right) \;,
\qquad  -\frac{\pi}{2\sqrt{\beta}} < \rho < \frac{\pi}{2\sqrt{\beta}} \;.
\end{equation}
The resulting equation now takes the form:
\begin{equation}
\left[ \frac{d^2}{d\xi^2}
       +2\,\lambda\,\delta\,\frac{s}{c}\,\frac{d}{d\xi}
       -\left\{\frac{1}{\lambda^2} - \lambda^2\delta(1+\delta)
        \right\}\frac{s^2}{c^2}
       +( \varepsilon + \lambda^2\delta )
\right]\Psi(\xi) = 0\;,
\end{equation}
where
\begin{equation}
\xi \equiv \frac{\rho}{\sqrt{m\hbar\omega}}\;,\quad
\lambda \equiv \sqrt{m\hbar\omega\beta}\;,\quad
\delta \equiv \frac{\gamma}{\beta}\;,\quad
\varepsilon \equiv \frac{2E}{\hbar\omega}\;,\quad
c \equiv \cos \lambda \xi\;,\quad
s \equiv \sin \lambda \xi\;.
\end{equation}
The equation resembles that for a particle trapped in a $\tan^2 x$ potential, 
which is much steeper than the harmonic oscillator potential.  
Using standard techniques, the energy eigenvalues are given by 
\begin{eqnarray}
E_n  & = & 
\hbar\omega\left\{ \left( n + \frac{1}{2}\right)
                   \sqrt{1 + \frac{\beta^2 m^2 \hbar^2 \omega^2}{4}}
                 + \left( n^2 + n + \frac{1}{2} \right) 
                   \frac{\beta m \hbar \omega}{2}
           \right\}\;,\cr
n & = & 0,\; 1,\;2,\; \ldots
\end{eqnarray}
The dependence on $n^2$ is reminiscent of the eigenvalues for a square well potential, 
and is reflective of the steeper rise of the transformed potential.  
Note that there is no dependence on the central parameter $\gamma$ in the eigenvalues. 
The eigenfunctions are Gegenbauer polynomials \cite{harmonic}.
The results are in agreement with those obtained by Kempf \cite{kempf}.

All of these can be generalized to $D$-dimensions.  
However, the non-commutativity of the position variables now plays a more important role, 
and it is not possible to handle general potentials in a straightforward manner.  
We begin by noting the momentum space representation for the position operator:
\begin{eqnarray}
\hat {x_i} & = & i\hbar\left\{ \left(1 + \beta p^2\right)\frac{\partial}{\partial p_i }
                                       + \beta' p_i p_j \frac{\partial}{\partial p_j} + \gamma p_i 
                       \right\} \;, \cr
\hat {p_i} & = & p_i \;.
\end{eqnarray}
Hermiticity requires
\begin{eqnarray}
\langle f | g \rangle & = & 
\int \frac{d^D \mathbf{p}}{\left[1 + (\beta + \beta') p^2 \right]^{1-\alpha}} \,
f^\ast(\mathbf{p}) \,g(\mathbf{p})\;, \cr
\alpha & = & \frac{ \gamma - \beta' \left(\dfrac{D-1}{2}\right) }
                  { (\beta + \beta') }\;.
\end{eqnarray}
For general potentials, the resultant Schr\"odinger equation is quite complicated.  
For central potentials, the existence of the angular momentum operator, 
and the resultant conservation enables us to separate variables, just as in the conventional case.  
\begin{eqnarray}
\sum_{i=1}^N \frac{\partial^2}{\partial p_i^2} 
& = & \frac{\partial^2}{\partial p^2}
    + \frac{D-1}{p} \frac{\partial}{\partial p}
    - \frac{L^2}{p^2} \;,\cr
\sum_{i=1}^N p_i \frac{\partial}{\partial p_i} 
& = & p\frac{\partial}{\partial p} \;,\cr
\phantom{\frac{1}{1}} 
L^2 & = & \ell (\ell + D -2 ), \qquad \ell = 0,1,2,\ldots
\end{eqnarray}
The radial equation can now be analyzed by adapting the methods described above.  
In this fashion, we obtain that in $D$-dimensions, the simple harmonic oscillator eigenvalues are given by
\begin{eqnarray}
\epsilon & \equiv & m\hbar\omega \;,\cr
E_{n\ell} 
& = & \hbar\omega
      \left[ \left(n + \frac{D}{2} \right)
             \sqrt{1 + \left\{\beta^2 L^2 + \frac{(D\beta + \beta')^2}{4}\right\}\epsilon^2} 
      \right.\cr
&   & \qquad\left.
    + \left\{ \left(\beta +\beta'\right)\left( n + \frac{D}{2} \right)^2
            + \left(\beta -\beta'\right)\left(L^2 + \frac{D^2}{4} \right) 
            + \beta'\frac{D}{2}
      \right\}\frac{\epsilon}{2}\;
      \right] \;,\cr
n & = & 0,\; 1, \; 2,\; \ldots
\end{eqnarray}
Once again, there is no dependence on $\gamma$. 
The corresponding eigenfunctions are given by Jacobi polynomials.  
Note that the accidental degeneracies are now lifted, and there is only the degeneracy 
associated with rotational symmetry.  We should  not be too surprised by this result 
of course, given that the classical orbits are not closed, 
and there are no Runge-Lenz vectors that we can define.  
In Ref.~\cite{harmonic}, we examined the effects of such deviations on oscillations of electrons in Penning traps.  
The effects are unfortunately quite small \cite{harmonic}.  

We might expect that a better bound on the minimal lengths and their associated 
effects might show up clearer in the many precision measurements associated 
with the Coulomb system.  Knowledge of the Coulomb system, in addition, 
could provide ways for how spin is incorporated in the presence of minimal lengths.

We have looked for the general solutions for the 3-D hydrogen atom, 
but have not been able yet to obtain closed expressions for the energy eigenvalues and eigenfunctions.   
The problem has to do with the proper definition of the operator ${\hat r}^{-1}$.  
In the process, we have solved for the eigenvalues for the ${\hat r}^2$ operator 
and its associated eigenfunctions.  By expanding the hydrogenic eigenfunctions in this basis, 
we have managed to solve for the eigenvalues and eigenfunctions numerically, 
with results that are consistent with the perturbative solutions obtained by Brau earlier \cite{brau}. 

I conclude by quoting the results for the eigenvalues for the ${\hat r}^2$ operator \cite{coulomb}:
\begin{eqnarray}
\frac{r_{nl}^2}{\hbar^2 (\beta + \beta')} 
& = & \left\{2n + \left(\frac{D}{2} + \ell 
                      + \sqrt{ \frac{\{(D-1)\eta + 1\}^2}{4} + \eta^2 L^2 }
                  \right)
      \right\}^2
    - (1-\eta)^2 \left\{ L^2 + \frac{(D-1)^2}{4}
                 \right\} 
\;,\cr
\eta & \equiv & {\beta\over \beta + \beta'}\;,
\qquad L^2 \;\equiv\; \ell(\ell + D - 2) \;.
\end{eqnarray}
The associated eigenfunctions are Jacobi polynomials.  
We can compare the numerical results for the energy eigenvalues with what is 
expected on the basis of perturbation theory for small values of $\beta$, $\beta'$.  
For the case of $\ell \neq 0$, and $D=3$, the perturbative shifts in energy are \cite{coulomb}: 
\begin{equation}
\Delta E_{n\ell} = \frac{\hbar^4}{m a^4 n^4}
\left[\, \left\{\, \frac{\beta - \beta'/2}{\ell(\ell+1)(\ell + 1/2)} 
             + \frac{2\beta + \beta'}{\ell + 1/2}
       \,\right\} n - \left( \beta + \beta'\right) 
\,\right]\;.
\end{equation}
Here, $m$ is the electron mass, and $a$ is the Bohr radius.  
It agrees very well indeed with the numerical results, and also agrees with an earlier 
result of Brau in the special case when $\beta = 2\beta'$.  
The formula contains a singularity when $\ell = 0$.  
The perturbation result is then critically dependent upon the form of the wavefunction at the origin.  
The minimal length hypothesis suggests that this behavior must be modified, 
perhaps with a cut-off on the integration consistent with the minimal eigenvalue for $\hat{r}^2$ noted above.  
The result is then consistent with the numerical results.  For further details, see Ref.~\cite{coulomb}.  

This work has been supported in part by a grant from the US Department of Energy, 
DE-FG05-92ER40709, Task A.

\end{document}